\documentclass[preprint,preprintnumbers,amsmath,amssymb,prb]{revtex4}
\usepackage{amsmath}
\usepackage{amsfonts}
\usepackage{graphicx}
\usepackage{color}
\usepackage{lscape}
\usepackage{amssymb}
\usepackage{threeparttable}
\usepackage{multirow}

\begin{document}

\title{Calculating excited states of molecular aggregates by the renormalized excitonic method}

\author{Yingjin Ma}
\author{Haibo Ma}
\email{haibo@nju.edu.cn}
\affiliation{Key Laboratory of Mesoscopic Chemistry of MOE, School of Chemistry and Chemical Engineering, Institute of Theoretical and Computational Chemistry, Nanjing University, Nanjing 210093, China}

\date{Latest revised on \today}

\begin{abstract}

In this paper, we apply the recently developed \emph{ab initio} renormalized excitonic method (REM) to the excitation energy calculations of various molecular aggregates, through the extension of REM to the time-dependent density functional theory (TDDFT).
Tested molecular aggregate systems include one-dimensional hydrogen-bonded water chains, ring crystals with $\pi$-$\pi$ stacking or van-der Waals interactions and the general aqueous systems with polar and non-polar solutes. The basis set factor as well as the effect of the exchange-correlation functionals are also investigated. The results indicate that the REM-TDDFT method with suitable basis set and exchange-correlation functionals can give good descriptions of excitation energies and excitation area for lowest electronic excitations in the molecular aggregate systems with economic computational costs. It's shown that the deviations of REM-TDDFT excitation energies from those by standard TDDFT are much less than 0.1 eV and the computational time can be reduced by one order.

\vspace{1ex}


\end{abstract}

\maketitle

\section{Introduction}

Molecular aggregates are coupled clusters of small molecules with intermolecular separations typically close to individual molecule size,
for example, the biological photosynthetic light harvesting system, the organic semiconductor crystal or the solute dissolved in solvents. Moreover, the ability of converting solar light into electrical or chemical energy in these systems through photosynthesis or photoelectric conversions motivates the study of the electronic excited states of molecular aggregates.
However, the theoretical characterization of these properties is often challenging to unravel due to their relatively large scales and complicated environments.
Among the current popular quantum chemical methods for calculating electronic excited states, time-dependent
density functional theory (TDDFT) is mostly widely used due to the good balance between the accuracy and computational cost \cite{Runge, TDDFTbook, Excited-Rev}.
It is well known that standard approximate exchange-correlation functionals used in DFT or TDDFT will underestimate the excitation energies for Rydberg states and charge-transfer states as well as extended $\pi$-conjugated systems and weakly interacting molecular aggregates. Such drawbacks are due to the fact that those functionals do not exhibit the correct $\frac{1}r$ asymptotic behavior and can not capture long-range correlation effects. Recent efforts have offered possibilities to account for long-range corrections and dispersion effects by the newly developed exchange-correlation functionals with long-range corrections \cite{LC1, LC2, LC3, LC4, LC5, LC6, LC7, LC8, LC9, LCx, LCx1, LCx2, LCx3, LCx4, LCx5} and/or dispersion corrections \cite{dispersion1, dispersion2, dispersion3, dispersion4, dispersion5, dispersion6}.
However, the applicable system size for excited state quantum chemistry calculation is still limited to a few hundred atoms at the most.  Since the $N^{3-4}$ scaling (N is the size of system) of the TDDFT \cite{TDDFTscale}, the application of TDDFT to very large systems is still challenging.

In order to reduce the computational scaling in TDDFT, many theoretical approaches \cite{Tretiak, ChenDFT, NOTDDFT, WernerCC, FCC1, FCC1add, FCC1add2, FCC2, LCC_eff, King, LCC2, LMRCIG, Marco, Shuai, WenjianTDDFT, adhoc, fde-tddft, Neugebauer07, Neugebauer10, Neugebauer08, Neugebauer10ct, Neugebauer11, Neugebauerbio} based on the local correlation approximation have been suggested. Chen and co-workers \cite{ChenDFT} developed a linear-scaling time-dependent density functional theory (TDDFT) algorithm using the localized density matrix (LDM) and in an orthogonal atomic orbital (OAO) representation.
Yang and co-workers \cite{NOTDDFT} extended this formalism and suggested reformulating TDDFT based on the non-orthogonal localized molecular orbitals (NOLMOs) \cite{NOLMO1, NOLMO2}. Casida and Wesolowski proposed the TDDFT within the frozen-density embedding (FDE) framework \cite{fde-tddft}, and Neugebauer and co-workers extended this approach with coupled electronic transtions \cite{Neugebauer07, Neugebauer10} and made applications of this approach to many interesting systems \cite{Neugebauer08, Neugebauer10ct, Neugebauer11, Neugebauerbio} like light-harvesting complexes in biomolecular assemblies \cite{Neugebauerbio}. Recently, based on the fragment LMOs that derived from capped fragments, Liu and co-workers \cite{WenjianTDDFT} suggested a new linear-scaling TDDFT method and successfully applied it to several large conjugated systems.

Considering the weak interactions between the molecular units, using a ``divide and conquer" idea to treat the excited states of the aggregated systems may be a worthwhile attempt \cite{fragREV}. In the fragment molecular orbital (FMO) method proposed by Kitaura and Fedorov \cite{FMObook, FMObook2}, the whole system can be divided into small fragments, and the total properties can be well estimated by the corresponding monomers, dimers, etc. Recently, using the FMO scheme, FMOx-TDDFT \cite{FMO1tddft, FMO2tddft, FMO3tddft, FMO4tddft} (x means n-body expansion) with analytic gradients have been developed and can give good descriptions for solvated and bio-chemical systems.
Mata and Stoll \cite{Stoll} also developed an improved incremental correlation approach for describing the excitation energies, with the inclusion of a dominant natural transition orbitals into selected excited fragment. However, these methods may lose efficiency when dealing with general systems which have multiple or uncertain excited regions.

For general systems, the Frenkel exciton model \cite{frenkel0} may be used as an alternative subsystem strategy and this model has been first applied to molecular crystals and subsequently extended to aggregates \cite{frenkel1, frenkel2, frenkel3, frenkel5, frenkelx, frenkelx1, frenkelx2, frenkel7, frenkel8, frenkel9, frenkel4}.
In its original form, the Frenkel exciton Hamiltonian describes a weakly interacting ensemble of two-level systems by
\begin{equation}\label{frenkel}
H=\sum_{i=1}\Omega_i b_i^+b_i+\sum_{i\neq j}s_{ij}(b_i^+b_j+b_j^+b_i)
\end{equation}
where indices \emph{i} and \emph{j} label ``blocks" (molecules), $b_i^+$ $(b_i)$ are the creation (annihilation) operators of an excitation on block $i$, $\Omega_i$ is the excited-state transition energy of block $i$, and $s_{ij}$ is the interaction, or coupling, between blocks $i$ and $j$.
The direct product of eigenstates of isolated blocks forms a convenient basis set for the global excited states. However, the non-diagonal couplings ($s_{ij}$) between different blocks, if calculated within the truncated Hilbert space directly, involve only the electrostatic interactions, and the description of excited states with such an approximation will fail for the systems in which the quantum dispersions dominate the inter-molecular interactions. Improvements have also been proposed through empirical corrections or exchange-correlation potentials \cite{frenkel7, frenkel8, frenkel9, frenkel4}.

In the recent years, the contractor renormalization group (CORE) method \cite{CORE, CORE1} and also the real-space renormalization group with effective interactions (RSRG-EI) \cite{RSRGEI} provided a novel kind of  subsystem methods for describing excited states of large systems, in which the general excited state is assumed to be an assembly of various block excitations (both single block excitations and multiple block excitations), and the interactions between adjoined blocks are taken into account through Bloch's effective Hamiltonian theory \cite{Bloch58, desC60}.
The basis set used in CORE or RSRG-EI is similar to that of Frenkel exciton model, but the non-diagonal coupling terms in CORE or RSRG-EI include not only the electrostatic interactions but also the quantum exchange contributions through the super-block calculation and the followed projection of the super-block wave-function onto the blocks' direct product basis, in contrast with those in traditional Frenkel exciton models which involve only the electrostatic interactions.
In 2005, Malrieu and co-workers \cite{REM05} made the further simplification, approximating the excited states of the whole system as only the linear combination of various single block excitations, and proposed it as the renormalized excitonic method (REM). Recently, we and our co-workers \cite{HJzhang, we2012} applied the REM into \emph{ab initio} quantum chemistry and successfully combined it with various \emph{ab initio} methods like full configuration interaction (FCI), configuration interaction singlet (CIS), and symmetry adapted cluster configuration interaction (SAC-CI). Good descriptions for excited states and ionized states of hydrogen chains and polyenes as well as polysilenes with economic computational costs have been achieved with both orthogonal localized molecular orbitals (OLMOs) and block canonical molecular orbitals (BCMOs) \cite{we2012}.

In this paper, we combine REM with TDDFT and extend the REM calculated systems from the linear molecule to various molecular aggregates.
Testing systems include hydrogen-bonded H$_2$O molecular chains, ring crystals like vdW interacted H$_2$O rings and $\pi$-$\pi$ stacked C$_2$H$_4$ rings, 2-D benzene aggregates, as well as aqueous systems with polar and non-polar solutes. The REM-TDDFT wavefunction as well as the basis set and functional factors are also discussed. The structure of this paper is designed as: In Sec. II, technical details of the method are introduced; in Sec III, we present calculated results of various molecular aggregates using REM-TDDFT and make comparisons with standard TDDFT calculations; and finally, we summarize and conclude our results in Sec. IV.

\section{Method}

\begin{figure}[htp]
\centering
\includegraphics[scale=0.45]{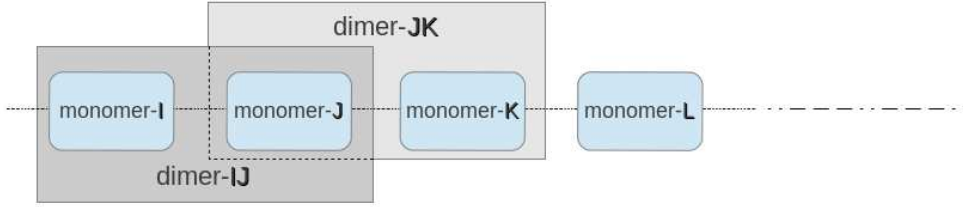}
\caption{The partition of the whole system into various blocks}
\label{fig-rem}
\end{figure}

The REM method is a type of fragment-based method. In REM, the whole system can be divided into many blocks (usually tens or hundreds of), as illustrated in Fig.~\ref{fig-rem}. Here the I, J, K, L are the block-monomers just like in Frenkel exciton model, and additionally, the adjacent monomers form the dimers.

In our previous work \cite{we2012}, we gave a detailed description of constructing REM Hamiltonian with BCMOs. Under BCMOs, the orthogonality only exists in the intra-blocks orbitals but not in the inter-blocks orbitals, and we used an approximate projector ({\emph{\~P}}$_0$) to unravel the non-orthogonality situation \cite{we2012}. Since this strategy is universal, we can use the subsystems' Kohn-Sham molecular orbitals (KS-MOs) instead of the BCMOs, and here we briefly introduce our REM-TDDFT strategy as below (up to 2-body interactions). For more details about the method, we refer the readers to our previous work \cite{we2012}.

1) Calculate each block-monomer by TDDFT to get KS-MOs, as well as the eigenstates for ground state ($\psi^0$) and an excited state ($\psi^{*}$). In our REM-TDDFT strategy, we assume the $\psi^0$ state is closed-shell ground state, and $\psi^{*}$ state is constructed by the excitation components in TDDFT. Then the basis functions ($|\Psi_{REM}\rangle $) of the model space are formed by
\begin{equation}\label{M8}
|\Psi_{REM} \rangle= \sum_{I=1}^{N} |\Psi_{I}^{*}\rangle =\sum_{I=1}^{N} [|\psi_I^{*}\rangle\prod_{J\neq I}|\psi_{J}^{0}\rangle]
\end{equation}
where $N$ is the total number of block monomers and the $I$, $J$ means the $I$-th, $J$-th block-monomer, respectively.
The corresponding projector ({\emph{\~P}}$_0$) by
\begin{equation}\label{projector}
\begin{split}
\tilde P_0 = |\Psi_{REM} \rangle S_m^{-1} \langle \Psi_{REM} | =  \sum_{I=1}^N  |\Psi_{I}^* \rangle (S_m^{-1})_{II'} \sum_{I'=1}^N\langle \Psi_{I'}^* | \\
 = \sum_{I=1}^{N} [|\psi_I^{*}\rangle\prod_{J\neq I}|\psi_{J}^{0}\rangle] (S_m^{-1})_{II'} \sum_{I'=1}^{N} [\langle\psi_{I'}^{*}|\prod_{J'\neq I'}\langle\psi_{J'}^{0}|] 
\end{split}
\end{equation}
where $S_m$ is the overlap matrix between model space basis functions.

Here we should mention that the antisymmetric property is satisfied in this product basis (Eq.~\ref{M8}). Although the MOs (or the electron density) of each monomer is totally localized on its own, quantum mechanics (QM) calculations of the dimers (or trimer, etc.) can redistribute the electron density \cite{FMObook}, which is crucial to discribe the charge transfer type excitation. Owing to the multi-mer's contribution, REM can give reasonable descriptions for various types of the low-lying excitations. 

2) Obtain two lowest excited states of dimer ($\psi_{IJ}^{*1}$ and $\psi_{IJ}^{*2}$) and use these eigenstates to form target states ($|\Psi_{IJ}^*\rangle$)
\begin{equation}\label{Mdim}
|\Psi_{IJ}^* \rangle=  |\psi_{IJ}^{*1}\rangle\prod_{\substack{ K\neq I,J} }|\psi_{K}^{0}\rangle+|\psi_{IJ}^{*2}\rangle\prod_{\substack{ K\neq I,J} }|\psi_{K}^{0}\rangle
\end{equation}

Following with projecting the target states onto model space via projector ({\emph{\~P}}$_0$)

\begin{equation}\label{project-cluster}
\begin{split}
\tilde P_0 | \Psi_{IJ}^* \rangle=  \sum_{I=1}^{N}|\Psi_I^{*} \rangle(S_m^{-1})_{II'} \sum_{I'=1}^N\langle \Psi_{I'}^* |  \Psi_{IJ}^* \rangle \\
= [\sum_{I'=1}^N(S_m^{-1})_{II'} \langle \Psi_{I'}^* |  \Psi_{IJ}^* \rangle ] \sum_{I=1}^{N}|\Psi_I^{*} \rangle
\end{split}
\end{equation}

Here we can denote the $[\sum_{I'=1}^N(S_m^{-1})_{II'} \langle \Psi_{I'}^* |  \Psi_{IJ}^* \rangle ]$ as matrix $C_0$. How to solve $C_0$ may be the most complicated part in the \emph{ab initio} REM strategy, one can refer to the Ref.~\cite{we2012} for detailed illustration. Nevertheless, we should also mention that two excited states are chosen here, as a result of only one excited state is kept in each monomer; if one more excited state were kept in monomer, then four excited states should be appropriate.

3) Once we get the $C_0$ matrix, usually the orthogonalization process should be applied to $C_0$ to get a new set of coefficients $C$, in order to obtain the Hermitian Hamiltonian. Then the expression of dimer-$IJ$' effective Hamiltonian can be written in the matrix form
\begin{equation}\label{eff-cluster}
H_{IJ}^{eff}=(C^+)^{-1}\varepsilon_{IJ}C^{-1}
\end{equation}
where the $\varepsilon_{IJ}$ are the two lowest excited states energies of dimer-${IJ}$. And the interactions between monomer-I and monomer-J can be acquired by
 \begin{equation}\label{heee}
{H}^{eff}_{I,J}={H}^{eff}_{IJ}-({H}^{eff}_{I}+{H}^{eff}_{J})
\end{equation}

\begin{figure}[htp]
\centering
\includegraphics[scale=0.5]{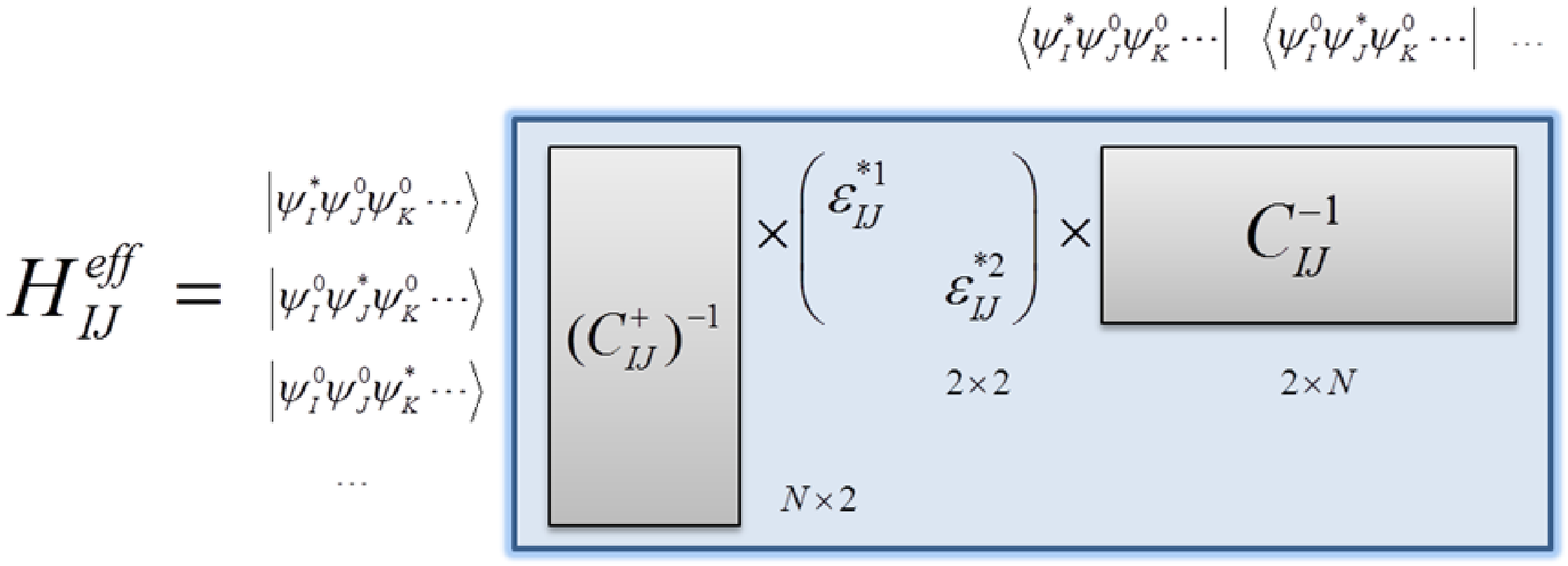}
\caption{Schematic illustration of the construction of $H_{IJ}^{eff}$}
\label{fig-Hij}
\end{figure}

Nevertheless, we should mentioned that the dimensions of various effective Hamiltonians are the same, and they are all equivalent to the number of REM basis. Here we take the $H_{IJ}^{eff}$ (Eq.~\ref{eff-cluster}) as example, as shown in Fig.~\ref{fig-Hij}. It could be found that the construction of $H_{IJ}^{eff}$ is under the whole REM bases while only two lowest excited states energies $\varepsilon_{IJ}^{*1}$ and $\varepsilon_{IJ}^{*2}$ are used, then the $H_{IJ}^{eff}$ will be a $N\times N$ matrix. When turn to $(H_{I}^{eff}+H_{J}^{eff}$), the $(\varepsilon_{I}^*+\varepsilon_{J}^0)$ and $(\varepsilon_{I}^0+\varepsilon_{J}^*)$ are used instead of $\varepsilon_{IJ}^{*1}$ and $\varepsilon_{IJ}^{*2}$, and the dimension is also $N\times N$.

4) After the determination of the effective Hamiltonians for various dimers, the renormalized Hamiltonian for the whole system can be obtained according to the following expressions
\begin{equation}\label{tth}
  {H}^{eff}=\sum_{I=1}^N {H}^{eff}_{I}+\sum_{I>J} {H}^{eff}_{I,J}
  \end{equation}
and finally can be solved as the generalized eigenvalue problem,
\begin{equation}\label{eff-final}
H^{eff}C^{eff}=S_mC^{eff}E
\end{equation}
where the eigenvalues $E$ are the excited state energies, and eigenstates $C^{eff}$ corresponds to contributions that the excitations occur in every blocks. In order to get the excitation energies, additional step should be applied to subtract the ground state energy with 2-body expansion ($E-E_0$), where
\begin{equation}\label{E_E0}
  E_0=\sum_{I}^{N}E_I^0+\sum_{I>J}(E_{IJ}^0-E_I^0-E_J^0)
\end{equation}
Since the dimension of the ${H}^{eff}$ is equivalent to the number of REM basis, which is usually less than one thousand, the Jacobi Method can be used to diagonalize the ${H}^{eff}$ of the whole system.

\section{Computational details}

We use our own code to implement the REM strategy. The preliminary TDDFT calculations on various blocks are implemented by GAUSSIAN\cite{g09} or GAMESS\cite{GAMESS}. In the 1-D H$_2$O chains systems, we use GAUSSIAN09 to do the subsystem TDDFT calculations on the monomers and dimers. In this test, both of the adjacent and the separate dimers are considered. In the other systems, if no extra illustration, the modified FMO subroutine in the GAMESS package is used to automatically select dimers and do the subsystem calculations. In our calculations the electrostatic potential (ESP) terms \cite{FMObook, FMObook2, FMOESP, ESPmore} and other correction terms are currently not considered, meaning only the original TDDFT calculations are implemented.

\section{Results and discussion}

\textbf{A. 1-D H$_2$O chains}

\begin{figure}[htp]
\centering
\includegraphics[scale=0.5]{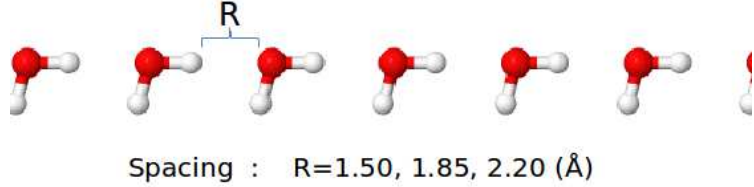}
\caption{The 1-D water molecular aggregates}
\label{fig-1d}
\end{figure}

The model 1-dimensional water molecular aggregates are chosen as the starting test systems.
The geometrical configurations are represented in Fig.~\ref{fig-1d}.  There the typical hydrogen bond length (1.85\AA) \cite{SCW_haibo} and other two spacings (1.50\AA \ and 2.20\AA) are chosen. The O-H bond length is fixed as 0.9584\AA, and the angle of H-O-H is fixed as 104.45$^\circ$. This starting system is simple and clear, and it should be a ideal model when we implement our starting REM-TDDFT calculation and do detailed analyses.

First, long-range corrected exchange-correlation functional LC-BLYP and Pople's 6-31+G* basis functions are used here to perform the REM-TDDFT calculations.
And in order to estimate the accuracy of REM-TDDFT, the standard TDDFT calculations are performed by GAUSSIAN09 \cite{g09}.
When performing the REM-TDDFT, two different fragmentation schemes are used here: fragmentation-A with one water molecule as one monomer, two water molecules as one dimer; fragmentation-B with two water molecules as one monomer, and four water molecules as one dimer. In fragmentation-A, each monomer keeps the ground state and one excited state. The excited state can be $S_1$ or $T_1$, depends on which state you want to calculate (of the whole system). The dimers here keep the lowest $S_1$ and $S_2$ (or $T_1$ and $T_2$) states. In fragmentation-B, the monomers and dimers contain double number of water molecules comparing to those in fragmentation-A, then there monomers keep one more excited state ($S_2$ or $T_2$), and dimers keep two more excited states ($S_3$, $S_4$ or $T_3$, $T_4$). All the electronic structure calculations on various monomers and dimers are also implemented by GAUSSIAN09.

\begin{table}[!hbp]
 \centering\scriptsize
 \begin{threeparttable}
 \caption{\label{tab-1D} Calculated singlet/triplet excitation energies (in eV) by standard TDDFT with LC-BLYP/6-31+G* and related excitation energy differences between REM-TDDFT and them}
  \begin{tabular}{lcccccccccccccccc}
  \hline
  \hline
  \multirow{2}{*}{System} & & \multicolumn{3}{c}{S$_1$ } & & \multicolumn{3}{c}{S$_2$ } & & \multicolumn{3}{c}{T$_1$ } & & \multicolumn{3}{c}{T$_2$ }\\
  \cline{3-5}  \cline{7-9} \cline{11-13} \cline{15-17}
          & & TDDFT & REM$^A$ & REM$^B$ &  & TDDFT & REM$^A$  & REM$^B$ & &
              TDDFT & REM$^A$ & REM$^B$ &  & TDDFT & REM$^A$  & REM$^B$  \\
  \hline
  $2.20$\AA    &        &        &        &        &        &        \\

  (H$_2$O$)_8$	 & & 7.942 & +0.018	& +0.024 & & 7.958 & +0.017	& +0.024 & & 7.198	& +0.021	&	+0.018	& &	7.212	&	+0.033	&	+0.013	   \\
 (H$_2$O$)_{16}$	 & &	7.930	&	+0.030	&	+0.036	& &	7.936	& +0.033	&	+0.046	
	 & &	7.188	&	+0.031	&	+0.028	& &	7.192	& +0.050	& +0.027	 \\
 (H$_2$O$)_{24}$	 & &	7.928	& +0.032	&	+0.038	& &	7.931	&	+0.038	&	+0.051	
	 & &	7.185	&	+0.034	&	+0.031	& &	7.188	&	+0.053	&	+0.030	  \\

  \hline
  $1.85$\AA      &          &          &          &          &          &     \\

  (H$_2$O$)_8$	 & &	7.899	&	-0.012	&	-0.004	& &	8.000	&	+0.092	&	+0.073	
	 & &	7.192	&	0.000	&	0.000	& &	7.284	&	+0.129	&	+0.033	 \\
 (H$_2$O$)_{16}$	 & &	7.899	&	-0.012	&	-0.004	& &	7.978	&	+0.111	&	+0.094	
	 & &	7.191	&	+0.001	&	+0.001	& &	7.263	&	+0.148	&	+0.049	 \\
 (H$_2$O$)_{24}$	 & &	7.899	&	-0.012	&	-0.004	& &	7.973	&	+0.116	&	+0.099	
	 & &	7.191	&	+0.001	&	+0.001	& &	7.258	&	+0.153	&	+0.053	 \\

  \hline
  $1.50$\AA      &          &          &          &          &          &     \\

  (H$_2$O$)_8$	 & &	7.575	&	-0.004	&	-0.016	& &	7.924	&	+0.048	&	+0.020	
	 & &	6.955	&	+0.029	&	-0.002	& &	7.288	&	+0.169	&	+0.035	 \\
 (H$_2$O$)_{16}$	 & &	7.576	&	-0.005	&	-0.017	& &	7.890	&	+0.079	&	+0.035	
	 & &	6.955	&	+0.029	&	-0.002	& &	7.252	&	+0.196	&	+0.057	 \\
 (H$_2$O$)_{24}$	 & &	7.577	&	-0.006	&	-0.018	& &	7.883	&	+0.083	&	+0.039	
	 & &	6.955	&	+0.029	&	-0.002	& &	7.245	&	+0.201	&	+0.062	 \\

  \hline
  \hline
  \end{tabular}
 \begin{tablenotes}
 \item[A] REM-TDDFT with H$_2$O as one monomer, (H$_2$O)$_2$ as one dimer
 \item[B] REM-TDDFT with (H$_2$O)$_2$ as one monomer, (H$_2$O)$_4$ as one dimer
  \end{tablenotes}
 \end{threeparttable}
\end{table}

\begin{figure}[!htp]
\centering
\includegraphics[scale=0.4]{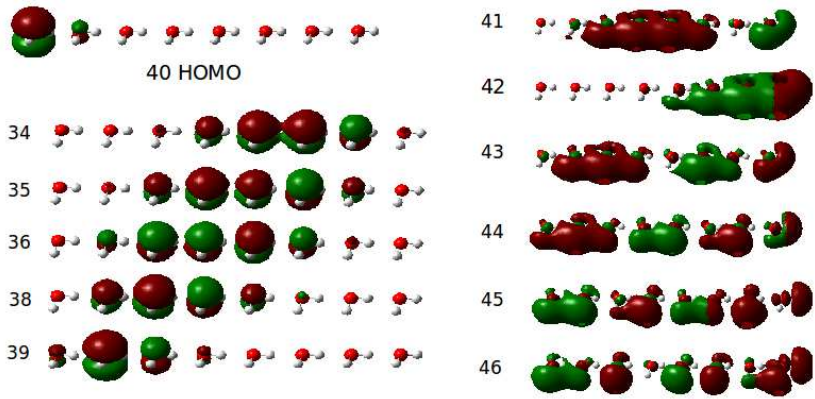}
\caption{Some important frontier orbitals in (H$_2$O)$_8$ chain}
\label{fig-h2o8}
\end{figure}

\begin{figure}[!htp]
\centering
\includegraphics[scale=0.5]{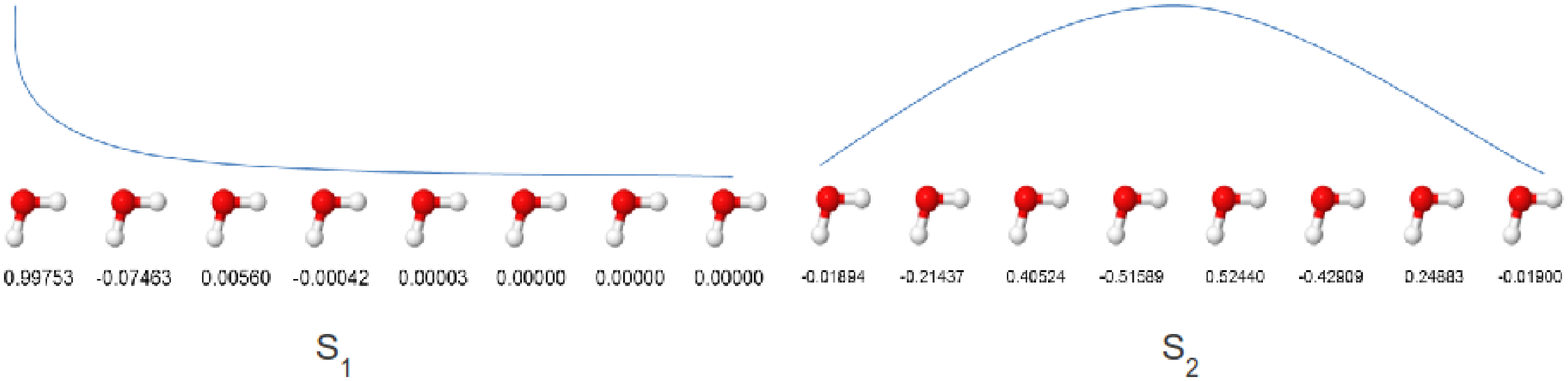}
\caption{Calculated contribution coefficients of various local excitations by REM-TDDFT for $S_1$ and $S_2$}
\label{fig-REM_wf}
\end{figure}

The REM-TDDFT results are summarized in Table.~\ref{tab-1D}. Let's start from the $S_1$ state and take the 1.85\AA \ spacing case (the typical hydrogen bond length) as an example. In REM-TDDFT with fragmentation-A, the standalone H$_2$O monomer's excitation energy for $S_1$ is 8.098 eV, and the standalone (H$_2$O)$_2$ dimer's excitation energy for $S_1$ is 7.887 eV. Here the results of REM-TDDFT in (H$_2$O)$_8$, (H$_2$O)$_{16}$ and (H$_2$O)$_{24}$ are all 7.887 eV for $S_1$ states. These values are agree with the full TDDFT quite well, with the error of about 0.01 eV. In REM-TDDFT with fragmentation-B, the standalone (H$_2$O)$_2$ monomer's excitation energy for $S_1$ is 7.887 eV, the (H$_2$O)$_4$ dimer's $S_1$ is 7.895 eV. There the corresponding REM results are all 7.895 eV in the different size of water aggregates, and match quite well with the full TDDFT values (7.899 eV). The $S_1$ states in 1.50\AA \ and 2.20\AA \ spacing cases have the behaviours similar with the 1.85\AA \ case, and the errors are all less than 0.04 eV. One may be confused about why the $S_1$ values in REM-TDDFT are not changed with the elongation of water chain. Here we take (H$_2$O)$_8$ for example to explain. From the result of full TDDFT calculation, some important orbitals are shown in Fig.~\ref{fig-h2o8} and some important configurations (excitation components) of $S_1$ are listed in Table.~\ref{tab-h2o8}. It could be found that, the excitation is mainly from the HOMO (40-th) orbital to various unoccupied orbitals. Combining with the Fig.~\ref{fig-h2o8}, it could be found that the electron mainly from left-most H$_2$O's $P_z$ orbital, excites to various unoccupied orbitals (no $P_z$ components) for $S_1$ state. Since the excitation mainly stems from the left-most water, the elongation of water chain will not change the $S_1$ excited energy much. We also show the $S_1$ wave functions obtained from the REM-TDDFT in the left part of Fig.~\ref{fig-REM_wf}. We could find from the coefficient analysis of REM wave function: the $S_1$ excitation is mainly contributed by the excitation from the left-most water, and a little component from the second left water. This picture is matching well with the full TDDFT calculation. Since there are only little components from the water monomers which are not belonging to the ``left two'', the excitation energy for the $S_1$ states will only slightly change with the increasing chain length.

\begin{table}[!hbp]
 \centering\footnotesize
 \begin{threeparttable}
 \caption{\label{tab-h2o8} Some important coefficients in the $S_1$ TDDFT amplitudes of (H$_2$O)$_8$}
  \begin{tabular}{lcccccccc}
  \hline
  \hline
      & Excitation mode & Amplitudes & & Excitation mode & Amplitudes & & Excitation mode  & Amplitudes \\
  \cline{2-3}  \cline{5-6} \cline{8-9}
  \cline{2-2}
  & 40 $\rightarrow$ 45 & -0.31473 & & 40 $\rightarrow$ 44 &  0.30290 & & 40 $\rightarrow$ 46 & -0.25920 \\
  & 40 $\rightarrow$ 43 & -0.19455 & & 40 $\rightarrow$ 47 & -0.19361 & & 39 $\rightarrow$ 43 & -0.12090 \\
  & 39 $\rightarrow$ 44 & -0.10897\\
  \hline
  \hline

  \end{tabular}
 \end{threeparttable}
\end{table}

\begin{table}[!hbp]
 \centering\footnotesize
 \begin{threeparttable}
 \caption{\label{tab-h2o8_s2} Some important coefficients in the $S_2$ TDDFT amplitudes of (H$_2$O)$_8$}
  \begin{tabular}{lcccccccc}
  \hline
  \hline
      & Excitation mode  & Amplitudes & & Excitation mode & Amplitudes & & Excitation mode & Amplitudes\\
  \cline{2-3}  \cline{5-6} \cline{8-9}
  & 38 $\rightarrow$ 42 &  0.31013 & & 36 $\rightarrow$ 42 &  0.29387 & & 35 $\rightarrow$ 43 & -0.17889 \\
  & 36 $\rightarrow$ 41 & -0.16899 & & 35 $\rightarrow$ 41 & -0.16083 & & 36 $\rightarrow$ 44 & -0.15601 \\
  & 38 $\rightarrow$ 43 &  0.15096 & & 35 $\rightarrow$ 45 & -0.12300 & & 34 $\rightarrow$ 41 &  0.11775 \\
  & 34 $\rightarrow$ 44 & -0.11517 & & 34 $\rightarrow$ 48 &  0.11011 & & \\
  \hline
  \hline
  \end{tabular}
 \end{threeparttable}
\end{table}

REM calculations for the triplet excited states usually give less error than those for the singlets,
and the triplet excitations usually have a more localized character than the singlet excitations \cite{we2012}. In Table.~\ref{tab-1D}, we can find the numerical accuracies for $T_1$ states are as well as those of $S_1$ states. In 1.85\AA \ spacing cases, the errors are only about 0.001 eV, no matter what fragmentation schemes we use. In 1.50\AA \ spacing, fragmentation-A has errors of about 0.03 eV. When enlarging the fragment units, the errors reduce to 0.002 eV. The errors increase to around 0.031 eV when the spacing turns to 2.20\AA \ (similar in $S_1$ case). This is due to the fact that the near-degeneracy problem for the very weak coupled systems will decrease the accuracy of REM calculations \cite{HJzhang}.

When turn to higher excited states, one could find the REM-TDDFT method can also give relative good descriptions. Here we take the $S_2$ state in 1.85\AA \ separation case as an example: In REM-TDDFT with fragmentation-A, the deviations between REM-TDDFT and full TDDFT are 0.092 eV, 0.111 eV and 0.116 eV for (H$_2$O)$_{8}$, (H$_2$O)$_{16}$, (H$_2$O)$_{24}$, respectively. These deviations are larger than those in $S_1$ states (about 0.01 eV) case. These deviations will turn to small when using larger monomers and dimers. With fragmentation-B, they are 0.073 eV, 0.094 eV and 0.099 eV, but still larger than those in $S_1$ case. One can also find the same trends in other spacing cases or in $T_2$ excited states. Why are the errors in $S_2$ ($T_2$) larger than those in $S_1$ ($T_1$)? In general, the accuracy of higher states depends on the lower states during the diagonalization process, therefore the error in the latter promotes a larger error in the former \cite{FMO2tddft}. There we can also use wave function analysis to illustrate this phenomenon in the REM-TDDFT calculation. In Table.~\ref{tab-h2o8_s2}, we list some important (largest) excitation components in $S_2$ of (H$_2$O)$_8$. It could be found that excitation in $S_2$ can originate from many orbitals (34, 35, 36 and 38), means this excitation may be contributed by many H$_2$O monomers. In REM strategy, we only consider the monomers and dimers, this hierarchical structure will lessen the change of domains on excitation and also affect the accuracy \cite{we2012}. In the previous $S_1$ state, the excitation mainly located on the edge, then the lessened change of domains may not affect the result much. However, in $S_2$ state, the excitation range from more H$_2$O units, then the lessened change of domains will affect the result more than in the $S_1$ state. Generally speaking, the higher excited states would own more excited regions, then the results will have larger errors. There we also show the REM-TDDFT $S_2$ wave function in the right part of Fig.~\ref{fig-REM_wf}. 
It could be found that our REM-TDDFT gives a normal distribution wave function with the peak in the middle of water chain. This picture agrees well with the full TDDFT calculation.

\textbf{B. Ring molecular crystals}

Now, let's turn to ring molecular crystals. Here we choose H$_2$O ring crystals dominated by vdW interacting and C$_2$H$_4$ $\pi$-$\pi$ stacked ring crystals to test. These systems can be seen as simplified systems with periodic boundary conditions. In this part, we also check whether the error is affected by the basis set and the DFT functional. The geometries and the detailed descriptions of these two types of ring crystals are refer to Ref.\cite{giantSAC}, water is put anti-parallel and ethylene parallel to each other. The alternating arrangement of waters would be favourable from the dipole-dipole interaction between the water monomers. The inter-water distance is 3.0\AA, near the 3.26\AA \ in which the ground state is attractive and has a minimum reflecting on the dipole-dipole interaction \cite{giantSAC}. The inter-ethylene distance is 4.5\AA, near the 4.90\AA \ in which the inter-ethylene electron transferred state is attractive and has a minimum at around 4.90\AA. In this inter-ethylene distance, there are two types of excitations: one is $\pi$ $\rightarrow$ $\pi$* excitations within each monomer; the other is electron-transfer type $\pi$ $\rightarrow$ $\pi$* excitations between monomers \cite{giantSAC}.
Parts of them are shown in Fig.~\ref{fig-ring}.

\begin{figure}[!htp]
\centering
\includegraphics[scale=0.2]{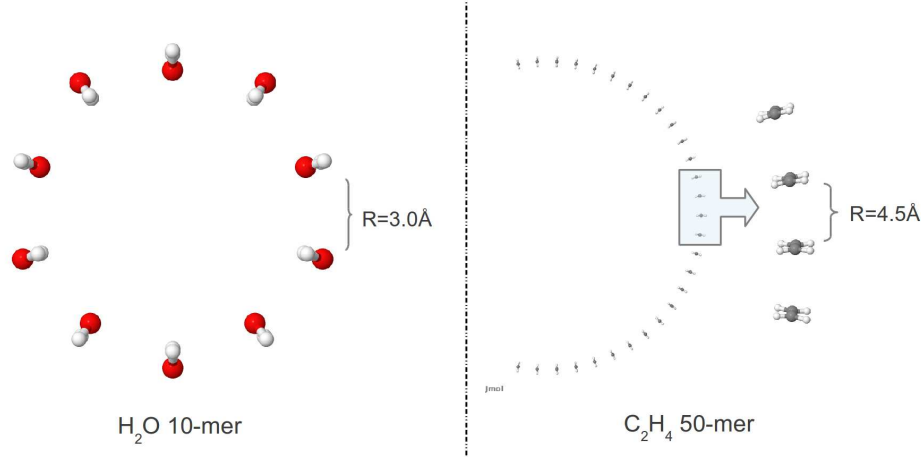}
\caption{The geometries of the two types of ring crystals}
\label{fig-ring}
\end{figure}

The long-range corrected functional (LC-BLYP) with three different basis functions (6-31G, 6-31+G* and 6-311++G**) are used here to perform the REM-TDDFT calculations and standard TDDFT. The subsystem TDDFT and standard TDDFT calculations are implemented by GAMESS \cite{GAMESS}. The results are listed in Table.~\ref{tab-ring}. In this table, we use two fragmentation schemes: the former (REM$^A$) is one H$_2$O (or C$_2$H$_4$) unit as one monomer, two H$_2$O (or C$_2$H$_4$) units as one dimer; the latter (REM$^B$) is two H$_2$O (C$_2$H$_4$) units as one monomer, then four H$_2$O (C$_2$H$_4$) units as one dimer. Each monomer keeps one ground state ($S_0$) and one excited state (S$_1$), each dimer keeps two lowest excited states ($S_1$, $S_2$). It could be found from the table that with former fragmentation scheme, the typical deviations in water ring systems between REM-TDDFT and standard TDDFT are about 0.06 eV in 6-31G, 0.14 eV in 6-31+G* and 0.11 eV in 6-311++G**, respectively. When using latter fragmentation scheme, those derivations turn to -0.01 eV, -0.06 eV and -0.07 eV, corresponding. It could be found that, no matter what fragmentation scheme is chosen, the errors for the larger basis sets are only slightly larger than the smaller basis sets on the average. The similar tendency can also be found in the ethylene ring systems. In general, the subs\begin{figure}[!htp]
\centering
\includegraphics[scale=0.2]{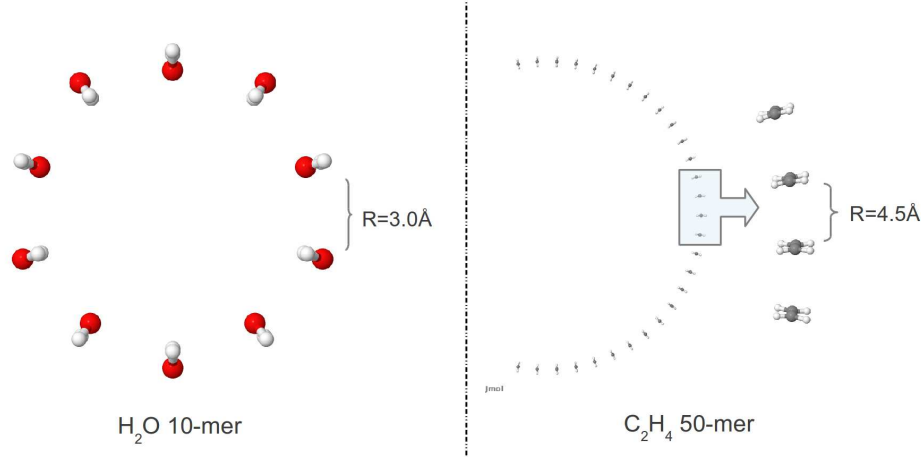}
\caption{The geometries of the two types of ring crystals}
\label{fig-ring}
\end{figure}ystem methods have a somewhat larger errors with extensive basis sets for the excitation energy. This is because the interactions betweens fragments will be enforced in extensive basis sets, such as the exchange-repulsion and charge transfer, however, there the REM-TDDFT method can only recover the interactions from two body level, it isn't enough.

\begin{table}[!hbp]
 \centering\small
 \begin{threeparttable}
 \caption{\label{tab-ring} Calculated $S_1$ excitation energies (in eV) by standard TDDFT and related excitation energy differences between REM-TDDFT results and them with different basis sets in the ring molecular crystals}
  \begin{tabular}{lcccccccccccccccc}
  \hline
  \hline
  \multirow{2}{*}{System} & & \multicolumn{3}{c}{LC-BLYP/6-31G} & & \multicolumn{3}{c}{LC-BLYP/6-31+G*} & & \multicolumn{3}{c}{LC-BLYP/6-311++G**} & \\
  \cline{3-5}  \cline{7-9} \cline{11-13} 
     &&  TDDFT & REM$^A$ & REM$^B$ & & TDDFT &  REM$^A$ & REM$^B$ && TDDFT & REM$^A$ & REM$^B$   \\
  \hline
  H$_2$O ring    &        &        &        &        &        &        \\
 (H$_2$O$)_{10}$ &&	6.763 &	+0.056 & -0.003	& &	6.840 &	+0.076 & -0.067	& &	5.857 &	+0.078 & -0.069	 &  \\
 (H$_2$O$)_{20}$ &&	6.758 &	+0.066 & -0.045	& &	6.837 &	+0.153 & -0.054	& &	5.841 &	+0.113 & -0.081	 &  \\
 (H$_2$O$)_{50}$ &&	6.756 &	+0.065 & -0.005	& &	6.845 &	+0.144 & -0.065	& &	5.836 &	+0.111 & -0.084	 &  \\

  \hline
 C$_2$H$_4$ ring &       &         &        &        &        &     \\

 (C$_2$H$_4$)$_{10}$ &&	8.132 &	+0.007 & -0.042	& &	7.108 &	-0.022 & -0.071	& &	6.886  & -0.010	& -0.060 &  \\
 (C$_2$H$_4$)$_{20}$ &&	8.119 &	+0.007 & -0.043	& &	7.080 &	-0.031 & -0.075	& &	6.843  & +0.022	& -0.024 &  \\
 (C$_2$H$_4$)$_{50}$ &&	8.117 &	+0.005 & -0.046	& &	7.074 &	-0.034 & -0.080 & & 6.845  & +0.018	& -0.027 &  \\

  \hline
  \hline
  \end{tabular}
\begin{tablenotes}
\item[A] REM-TDDFT with H$_2$O or C$_2$H$_4$ as one monomer
\item[B] REM-TDDFT with (H$_2$O)$_2$ or (C$_2$H$_4$)$_2$ as one monomer
\end{tablenotes}
 \end{threeparttable}
\end{table}

Next, it is also of interest to see if the error is affected by the DFT functional.
There we compare the $S_1$ excitation energies using BLYP, B3LYP with 6-31+G* basis, and also add the LC-BLYP data in Table.~\ref{tab-ring}. There we also use the two fragmentation schemes as above. The results are summarized in Table.~\ref{tab-ring2}. We can find that the REM with long-range corrected functional LC-BLYP give a better description than the pure functional BLYP and the hybrid functional B3LYP. When using BLYP or B3LYP functional, the typical deviation in REM$^A$ is about 0.3 eV in water crystals, and even larger than 0.5 eV in ethylene crystals (even 1.0 eV in (C$_2$H$_4$)$_{10}$). These deviations can be decreased using larger fragmentation scheme: in REM$^B$, the typical deviations in water crystals decrease from 0.3 eV to about 0.15 eV, and in ethylene crystals, the typical deviations can reduce to about 0.1 eV.
Obviously, such performances with BLYP and B3LYP are generally not satisfactory, since BLYP usually underestimate of non-local long-range electron-electron exchange interactions by the pure density functionals, and the B3LYP usually give wrong description of long-range interactions for this functional stem from modeling strong intermolecular interactions between solid macromolecular systems. This implies that the long-range corrections are very important for the REM-TDDFT calculations of large systems.

\begin{table}[!hbp]
 \centering\small
 \begin{threeparttable}
 \caption{\label{tab-ring2} Calculated $S_1$ excitation energies (in eV) by standard TDDFT and related excitation energy differences between REM-TDDFT results and them with different functionals in the ring molecular crystals }
  \begin{tabular}{lcccccccccccccccc}
  \hline
  \hline
  \multirow{2}{*}{System} & & \multicolumn{3}{c}{BLYP/6-31+G*} & & \multicolumn{3}{c}{B3LYP/6-31+G*} & & \multicolumn{3}{c}{LC-BLYP/6-31+G*} & \\
  \cline{3-5}  \cline{7-9} \cline{11-13} 
          & &  TDDFT & REM$^A$ & REM$^B$ & & TDDFT &  REM$^A$  & REM$^B$ & &  TDDFT &REM$^A$ & REM$^B$   \\
  \hline
  H$_2$O ring    &        &        &        &        &        &        \\

 (H$_2$O$)_{10}$      && 5.329 & -0.266	& -0.135 &&	6.593 &	+0.113	& -0.112 & & 6.840  & +0.076 &	-0.067	  \\
 (H$_2$O$)_{20}$      && 5.355 & +0.278	& -0.155 &&	6.626 &	+0.161	& -0.143 & & 6.837	& +0.153 &	-0.054	  \\
 (H$_2$O$)_{50}$      && 5.357 & +0.336	& -0.158 &&	6.643 &	+0.155	& -0.148 & & 6.845	& +0.144 &	-0.065	 \\

  \hline
   C$_2$H$_4$ ring &       &         &        &        &      &     \\

 (C$_2$H$_4$)$_{10}$  && 5.329 & -1.091 & -0.338  && 6.297 & -0.456	&	-0.135	& &	7.108	 &	-0.022	 &	-0.071 \\
 (C$_2$H$_4$)$_{20}$  && 5.056 & -0.847	& -0.098  && 6.243 & -0.445	&	-0.117	& &	7.080	 &	-0.031	 &	-0.075 \\
 (C$_2$H$_4$)$_{50}$  && 5.038 & -0.835	& -0.090  && 6.235 & -0.450	&	-0.120	& &	7.074	 &	-0.034	 &	-0.080 \\

  \hline
  \hline
  \end{tabular}
  \begin{tablenotes}
\item[A] REM-TDDFT with H$_2$O or C$_2$H$_4$ as one monomer
\item[B] REM-TDDFT with (H$_2$O)$_2$ or (C$_2$H$_4$)$_2$ as one monomer
  \end{tablenotes}
 \end{threeparttable}
\end{table}

\quad

\textbf{C. 2-D benzene crystal systems}

\begin{figure}[!htp]
\centering
\includegraphics[scale=0.30]{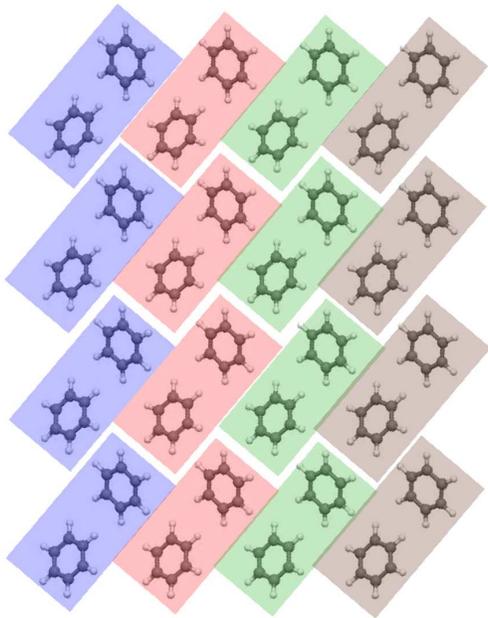}
\caption{The geometry of the 2-D benzene crystal. }
\label{fig-benzene}
\end{figure}

In this part, we attempt to apply the REM-TDDFT calculations to the 2-D benzene crystal system. It is well known that crystals of acenes such as pentacene and tetracene own great potentials in the organic photovoltaic field \cite{STchange} and consequently the accurate calculations of the electronic excited states of such aggregates are highly desired.
The benzene crystal system can be seen as a simple model system of the acene crystals.
Here we choose a number of benzene molecules in the (1,0,0) crystal face of the benzene crystal \cite{bencrystal} to do our test. As illustrated in Fig.~\ref{fig-benzene},
each benzene column own its unique colour and four columns make up the 2-D benzene aggregates. One color square means one monomer and the dimers are automatically selected by the FMO subroutine in the GAMESS package. The LC-BLYP functional with 6-31G basis sets are used here.
When doing the REM-TDDFT calculations, each monomer keeps one ground state ($S_0$) and one excited state ($S_1$) and each dimer keeps one ground state ($S_0$) and two excited states ($S_1$, $S_2$). 

The results of calculated $S_1$ excitation energies are listed in Table.~\ref{tab-2d}. It could be found that the performance of REM-TDDFT exists in the 1-column situation: the difference between the result of REM-TDDFT and TDDFT is only -0.003 eV. When the system tends to extend by adding the columns, the TDDFT results gradually converge to 5.672 eV. It means that the properties of the 2-D benzene system with such large sizes are already approaching the bulk ones of 2-D infinite benzene crystal. Here the results of REM-TDDFT are also converging very well to 5.660 eV, however, the difference for excitation energies of $S_1$ between REM-TDDFT and TDDFT increases to -0.012 eV, slightly larger than that in the 1-column case but still satisfactory. Such minor excitation energies errors with the magnitude from -0.003 eV to -0.012 eV for REM calculations of the 2-D benzene systems are comparable to those in the above 1-D examples, implying that REM has the potential to be applied to realistic molecular aggregates with complicated morphologies.

\begin{table}[!hbp]
 \centering\small
 \begin{threeparttable}
 \caption{\label{tab-2d}Calculated $S_1$ excitation energies (in eV) by REM-TDDFT and by standard TDDFT and the difference $\Delta$ (in eV) between the results of REM-TDDFT and TDDFT. The LC-BLYP/6-31G are used in the calculations.}
  \begin{tabular}{lccccccc}
  \hline
  \hline
 & System  & 1-Column & 2-Columns & 3-Columns & 4-Columns     \\
  \hline
 &REM-TDDFT &  5.682  & 5.664 & 5.660 & 5.660  \\
 & TDDFT    &  5.685  & 5.671 & 5.672 &  5.672 \\
 &  $\Delta$ & -0.003 & -0.007 & -0.012 & -0.012  \\
    \hline
  \hline
  \end{tabular}
 \end{threeparttable}
\end{table}

\textbf{D. Solvated systems}

Finally, let's turn our focus on the solutions. We choose two typical systems: one is the benzene + (H$_2$O)$_n$ system, the other is acetone + (H$_2$O)$_n$ system.
These represent different solvation behaviours of non-polar and polar solutes dissolved in water.
The geometries of the those two systems are chosen from the molecular dynamics (MD) trajectories in our other work \cite{weother}.
Firstly, we performing the NVT MD simulations with simple point charge extended (SPC/E) potential \cite{SPCE} for water and OPLS potential (optimized potentials for liquid simulations) \cite{OPLS0, OPLS} for benzene and acetone, then select five uncorrelated snapshots for each set. From the selected snapshots, 12, 36 or 60 water molecules closest to benzene (acetone) as well as the central benzene (acetone) are taken to perform the REM-TDDFT calculations.
Long-range corrected functional (LC-BLYP) with 6-31+G* basis functions are used here.
There we treat one molecular unit as one fragment-monomer, then two molecular units as one fragment-dimer. In this test, we choose all of the benzene-water dimers, and the water-water dimers with interval less than 3.0\AA. Each monomer keeps the ground state $S_0$ and the first singlet excited state $S_1$, each dimer keeps $S_0$, $S_1$ and also the second singlet excited state $S_2$.
The results of both REM-TDDFT and standard TDDFT of these two solvated systems are listed in Table.~\ref{tab-solvated1} and Table.~\ref{tab-solvated2}, separately. 

The results of solvated benzene are listed in Table.~\ref{tab-solvated1}. It could be found that the REM-TDDFT can reproduce the full TDDFT values quite accurately. The average deviation in the benzene + (H$_2$O)$_{12}$ systems is only -0.006 eV, with mean square error of 0.005 eV. The deviations will slightly increase with more water molecules: when adding 36 H$_2$Os, the average deviation increase to -0.014 eV, with mean square error of 0.011 eV; and when adding 60 H$_2$Os, these two values turn to -0.027 eV and 0.021 eV, respectively. The deviations turn to large with the enlarged systems, for there are much more many-body interactions in those enlarged systems. Here we can also observe that the REM-TDDFT excitation energies are underestimated frequently relative to the full TDDFT values, the similar phenomenon is also observed in FMO2-TDDFT calculations \cite{FMO2tddft}. Introducing the 3-body interactions could improve the results , since only the two body interactions usually gives negative pair corrections \cite{HJzhang, FMO3tddft}.

The results of solvated acetone are listed in Table.~\ref{tab-solvated2}. It could be found that the REM-TDDFT can also reproduce the full TDDFT values quite well. The average deviations and the mean square errors (in brackets) are -0.020 eV (0.011 eV), -0.018 eV (0.073 eV) and -0.008 eV (0.066 eV), correspondingly. These deviations are larger than those in the solvated benzene, for the polar acetone is soluble in water, and there are stronger interactions between solute and solvent than those in solvated benzene systems. In solvated acetone systems, the electrons can be excited from acetone to acetone itself, and from acetone to the neighbor waters. While in the solvated benzene systems, the $S_1$ excitations are mainly localized on the benzene molecule itself. In principle, one need to enlarge the fragment units or introducing 3-body (or higher) interactions to give a better description. Although lacking some interactions information, the wave functions of REM-TDDFT can also give qualitative correct pictures in these two solvated systems: the excitations are mainly from center benzene (or acetone) molecule (about 0.95-0.99, depends on the system); the contributions from waters are very small, and decrease with elongation of benzene-water (or acetone-water) distance.

\begin{table}[!hbp]
 \centering\small
 \begin{threeparttable}
 \caption{\label{tab-solvated1} Calculated $S_1$ excitation energies (in eV) in solvated benzene systems by standard TDDFT and related excitation energy differences between REM-TDDFT results and them}
  \begin{tabular}{lcccccccccccccccc}
  \hline
  \hline
  \multirow{2}{*}{System} & & \multicolumn{3}{c}{benzene+(H$_2$O)$_{12}$} & & \multicolumn{3}{c}{benzene+(H$_2$O)$_{36}$} & & \multicolumn{3}{c}{benzene+(H$_2$O)$_{60}$} & \\
  \cline{3-5}  \cline{7-9} \cline{11-13} 
          & &  REM & TDDFT & $\Delta^*$ & &  REM  & TDDFT & $\Delta^*$
          & &  REM & TDDFT & $\Delta^*$ \\
  \hline
 snapshot-1 && 5.349 & 5.359 &-0.010 && 5.322& 5.339& -0.017 && 5.317 & 5.340 &-0.023\\
 snapshot-2 && 5.309 & 5.307 & 0.002 && 5.304& 5.323& -0.019 && 5.298 & 5.355 &-0.057\\
 snapshot-3 && 5.313 & 5.317 &-0.004 && 5.262& 5.291& -0.029 && 5.263 & 5.296 &-0.033\\
 snapshot-4 && 5.281 & 5.286 &-0.005 && 5.240& 5.250& -0.010 && 5.195 & -$^\Diamond$  &  -   \\
 snapshot-5 && 5.268 & 5.280 &-0.012 && 5.272& 5.267&  0.005 && 5.283 & 5.280 & 0.003\\
  \hline
 $\overline{\Delta}^{(\sigma)}$ && & &-0.006$^{(0.005)}$ && & & -0.014$^{(0.011)}$ && & &  -0.027$^{(0.021)}$   \\
  \hline
  \hline
  \end{tabular}
 \begin{tablenotes}
 \item[*] The difference is the result of REM-TDDFT minus that of standard TDDFT
 \item[$\Diamond$] Accurate result is unavailable for the convergence problem in DFT
 \end{tablenotes}
 \end{threeparttable}
\end{table}

\begin{table}[!hbp]
 \centering\small
 \begin{threeparttable}
 \caption{\label{tab-solvated2} Calculated $S_1$ excitation energies (in eV) in solvated acetone systems by standard TDDFT and related excitation energy differences between REM-TDDFT results and them}
  \begin{tabular}{lcccccccccccccccc}
  \hline
  \hline
  \multirow{2}{*}{System} & & \multicolumn{3}{c}{acetone+(H$_2$O)$_{12}$} & & \multicolumn{3}{c}{acetone+(H$_2$O)$_{36}$} & & \multicolumn{3}{c}{acetone+(H$_2$O)$_{60}$} & \\
  \cline{3-5}  \cline{7-9} \cline{11-13} 
          & &  REM & TDDFT & $\Delta^*$ & &  REM  & TDDFT & $\Delta^*$
          & &  REM & TDDFT & $\Delta^*$ \\
  \hline
snapshot-1 && 4.218 & 4.229 & -0.011 && 4.245 & 4.274 & -0.029 && 4.232 & 4.290&-0.058\\
snapshot-2 && 4.195 & 4.210 & -0.015 && 4.196 & 4.227 & -0.031 && 4.192 & 4.229&-0.037\\
snapshot-3 && 4.311 & 4.345 & -0.034 && 4.477 & 4.354 &  0.123 && 4.488 & 4.382& 0.106\\
snapshot-4 && 4.341 & 4.374 & -0.033 && 4.363 & 4.430 & -0.067 && 4.247 & -$^{\Diamond}$ &  -   \\
snapshot-5 && 4.368 & 4.375 & -0.007 && 4.340 & 4.424 & -0.084 && 4.366 & 4.408&-0.042\\
  \hline
$\overline{\Delta}^{(\sigma)}$ && & & -0.020$^{(0.011)}$ && & & -0.018$^{(0.073)}$ && & & -0.008$^{(0.066)}$\\
  \hline
  \hline
  \end{tabular}
 \begin{tablenotes}
 \item[*] The difference is the result of REM-TDDFT minus that of standard TDDFT
 \item[$\Diamond$] Accurate result is unavailable for the convergence problem in DFT
 \end{tablenotes}
 \end{threeparttable}
\end{table}

At last, we briefly introduce the timings for REM-TDDFT method. The time costs of both REM-TDDFT and standard TDDFT calculations for testing systems are listed in Table.~\ref{tab-time}. All calculations are implemented by the Sugon 12-core servers with Intel Xeon X5650@2.67GHz. It can be found from the table that the REM-TDDFT costs less time than standard TDDFT in this server, for the REM-TDDFT strategy has the approximate $c_1\times {N_e}^3$ + n $\times c_2 \times {N_e}^3$ scaling when considering only the two-body interactions \cite{we2012}. The former is mainly from the calculation of the overlap matrix in the model space, and the latter is caused by the projections from target space to model space. There $n$ is the number of dimers and $N_e$ is the number of electrons of the whole system, while $c_1, c_2$ are constants affected by the preserved number of configurations in each state. Here the ${N_e}^3$ in mainly from the lower triangular-upper triangular (LU) decomposition \cite{we2012}, there the time-scale factor can at most up to ${N_e}^3$, in practical applications it can be at most reduced to ${N_e}^{2.376}$ \cite{n2.376}. In fact, since REM-TDDFT using a disentanglement way \cite{we2012, DMRGCORE} to get the two body interactions, the various interactions extracting from dimers can be easily distribute to many servers, then the time costs can be even lower.

\begin{table}[!hbp]
 \centering\small
 \begin{threeparttable}
 \caption{\label{tab-time}The approximate wall clock timing for REM-TDDFT and standard TDDFT calculations at LC-BLYP/6-31+G* level in the solvated acetone systems in Table.~\ref{tab-solvated2}.}
  \begin{tabular}{lccc}
  \hline
  \hline
 System  & REM-TDDFT*$^{,\diamond}$ & TDDFT $^\diamond$  \\
  \hline
  acetone + (H$_2$O)$_{12}$ &\quad $\sim$0.5 min \quad& \quad$\sim$8 min \quad \\
  acetone + (H$_2$O)$_{36}$ &\quad $\sim$11 min \quad& \quad$\sim$200 min \quad \\
  acetone + (H$_2$O)$_{60}$ &\quad $\sim$58 min \quad& \quad$\sim$530 min \quad \\
  \hline
  \hline
  \end{tabular}
 \begin{tablenotes}
 \item[*] The timings of calculations of the monomers and dimers are not counted in.
 \item[$\diamond$] 12-core server with Intel Xeon X5650
 \end{tablenotes}
 \end{threeparttable}
\end{table}

\quad



\section{Summary and conclusion}

In this paper, we extend the \emph{ab initio} REM method to TDDFT theory and use this approach to calculate electronic excitation energies of various molecular aggregates. It is shown that this approach can not only gives a description of electronic excitation energies, but also provides a qualitative picture where the excitation locates. Since only the subsystems need to be solved in the whole aggregates, the computational costs are reduced remarkably than the TDDFT calculations while losing only little accuracy. Such achievements provide a new promising sub-system methodology for future quantitative studies of large complicated systems such as supramolecules, condensed phase matters.

Test calculations for the one dimensional water molecule chains show that REM-TDDFT method is effective in reproducing the electronic excitation energies of low-lying excited states: the typical deviation is only about 0.030 eV in $S_1$ or $T_1$ states, and slight larger for higher excited states. The wave function analysis of REM-TDDFT also gives correct pictures of the excitation behavior in these systems. Furthermore, we test the REM-TDDFT with different basis sets and also various exchange-correlation functionals. We find that the larger basis sets will only slightly affect the final results, but the DFT functionals would significantly influence the stability and accuracy. Here the long-range corrected functionals with appropriate basis sets are recommended for dealing with large molecular aggregates.
The trial test on the 2-D structure like benzene aggregates are also implemented and satisfactory excitation energy accuracies are also observed for them.
At last, we turn to two types of aqueous systems to examine our REM-TDDFT's performances for the solutions. With LC-BLYP functional and 6-31+G* basis sets, our REM-TDDFT method can reproduce the standard TDDFT values quite well, for both of the aqueous systems with polar and non-polar solutes.

The results of REM-TDDFT are acceptable in these molecular aggregate systems, however, if one wants to pursue more accurate results the higher many-body interactions and ESP effects should be introduced. Progress along this direction is being made in our laboratory.

\section*{Acknowledgment}
This work is supported by the National Natural Science Foundation of China (Grant No. 21003072 and 91122019), National Basic Research Program (Grant No. 2011CB808604) and the Fundamental Research Funds for the Central Universities. We are grateful to C. G. Liu, Y. Liu, H. J. Zhang for the stimulating conversations.

\clearpage

\begin{figure}[!htp]
\centering
\includegraphics[scale=1.0]{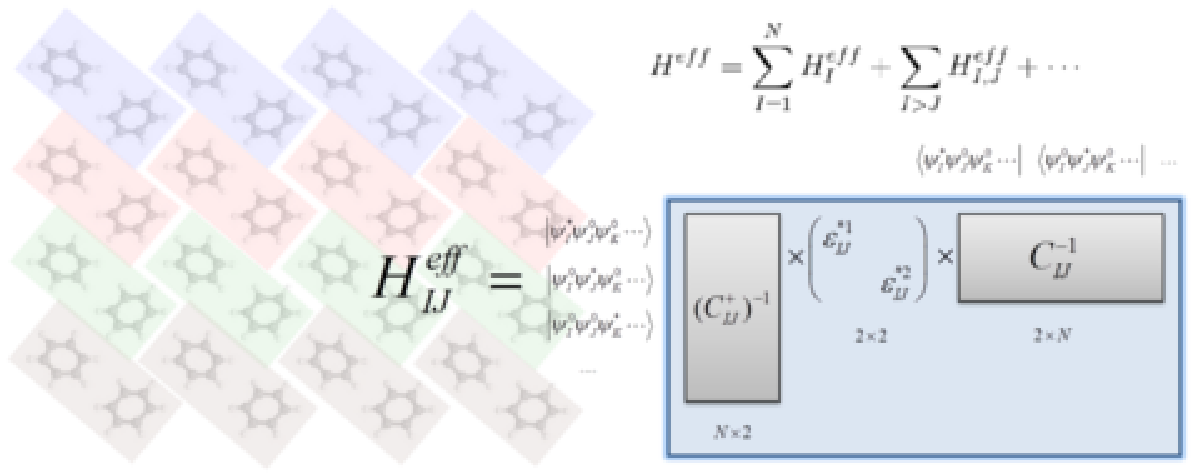}
\caption{TOC}
\end{figure}

\end{document}